\documentclass[10pt,twocolumn,letterpaper]{article}

\usepackage{cvpr}              

\usepackage{graphicx}
\usepackage{amsmath}
\usepackage{amssymb}
\usepackage{booktabs}
\usepackage{tikz}
\DeclareMathOperator*{\argminB}{argmin} 

\usepackage[pagebackref,breaklinks,colorlinks]{hyperref}

\usepackage[capitalize]{cleveref}
\crefname{section}{Sec.}{Secs.}
\Crefname{section}{Section}{Sections}
\Crefname{table}{Table}{Tables}
\crefname{table}{Tab.}{Tabs.}

\def\cvprPaperID{72} 
\def\confName{CVPR}
\def\confYear{2024}

\title{Simulation of a Vision Correction Display System}

\author{Vidya Sunil\\ 
Vehant Technologies\\
{\tt\small vidyas@vehant.com}
\and
Renu M Rameshan\\
Vehant Technologies\\
{\tt\small renur@vehant.com}
}

\begin{document}
\twocolumn[{%
\renewcommand\twocolumn[1][]{#1}%
\maketitle
\begin{center}
    \centering
    \captionsetup{type=figure}
    \begin{subfigure}{0.22\linewidth}
    \includegraphics[width=\linewidth]{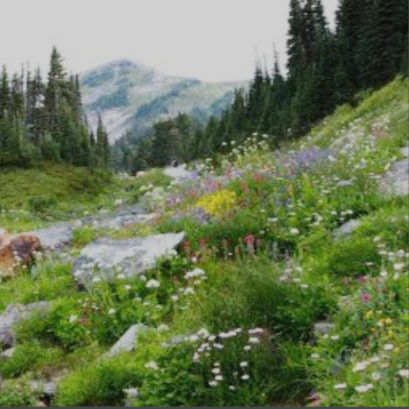}
    \caption{Original Image}
    \label{fig:a}
  \end{subfigure}
  \hfill
  \begin{subfigure}{0.22\linewidth}
    \includegraphics[width=\linewidth]{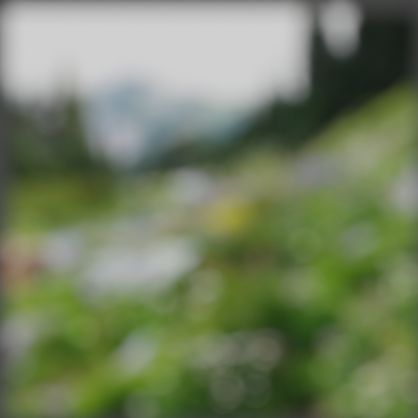}
    \caption{Defocused Image}
    \label{fig:b}
  \end{subfigure}
   \hfill
  \begin{subfigure}{0.22\linewidth}
    \includegraphics[width=\linewidth]{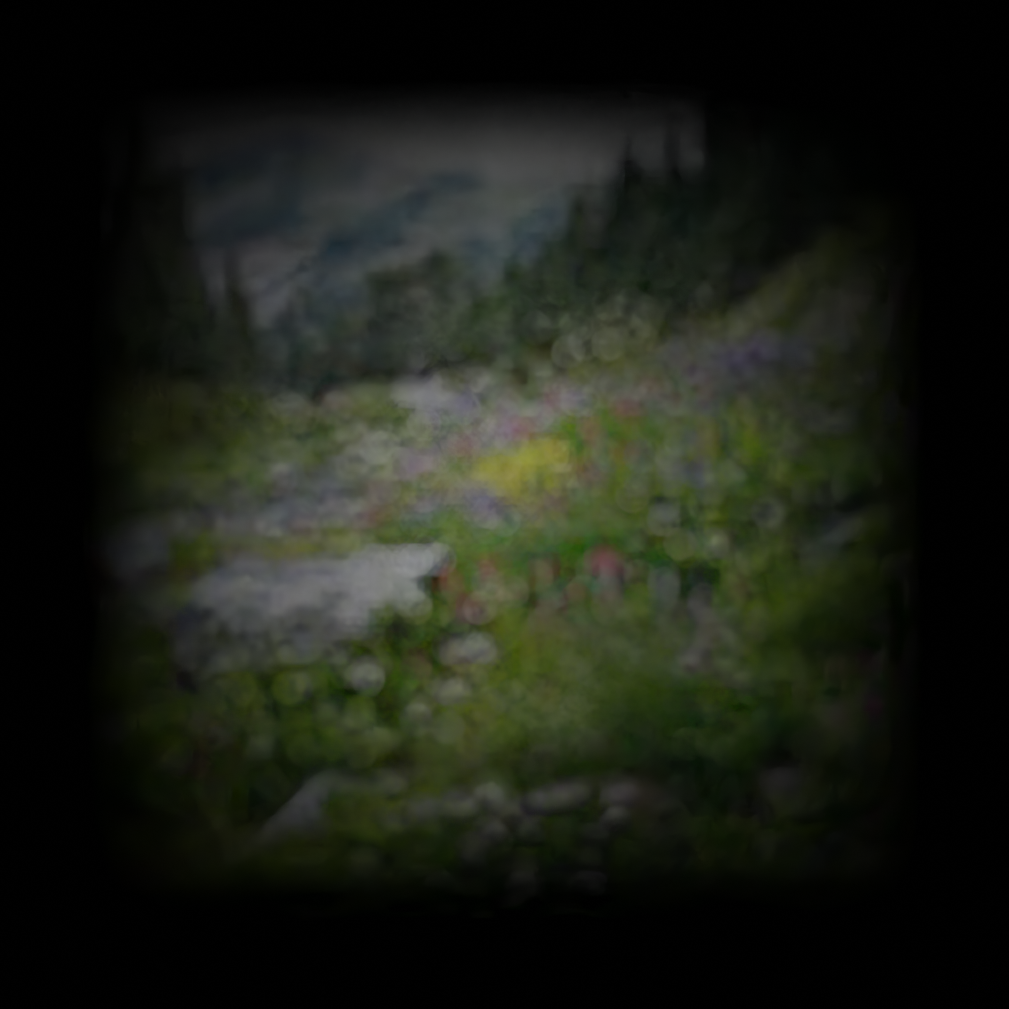}
    
    \caption{VCD image with pinhole array}
    \label{fig:c}
  \end{subfigure}
   \hfill
  \begin{subfigure}{0.22\linewidth}
    \includegraphics[width=\linewidth]{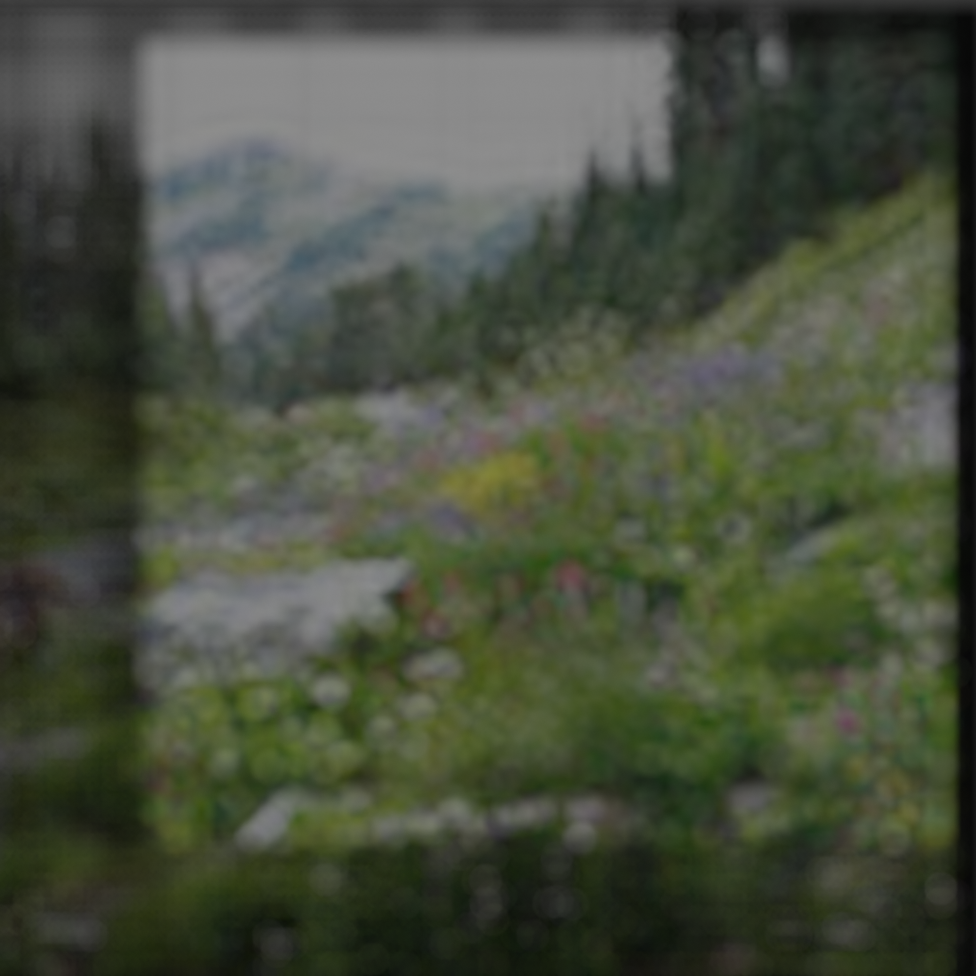}
    
    \caption{ VCD image with lenslet array}
    \label{fig:d}
  \end{subfigure}
  \caption{Output images (\ref{fig:c}, \ref{fig:d}) of a simulated  vision correction display for a hyperopic eye which sees \ref{fig:b} without any correction.}
  \label{fig:First}
\end{center}%
}]


\begin{abstract}
   Eyes serve as our primary sensory organs, responsible for processing up to 80\% of our sensory input. However, common visual aberrations like myopia and hyperopia affect a significant portion of the global population.
   This paper focuses on simulating a Vision Correction Display (VCD) to enhance the visual experience of individuals with various visual impairments. Utilising Blender, we digitally model the functionality of a VCD in correcting refractive errors such as myopia and hyperopia. With these simulations we can see potential improvements in visual acuity and comfort. These simulations provide valuable insights for the design and development of future VCD technologies, ultimately advancing accessibility and usability for individuals with visual challenges.
\end{abstract}

\section{Introduction}
\label{sec:intro}

Almost all optical systems, including the human eye, exhibit optical aberrations. In the past, eyeglasses were the primary means of correcting optical aberrations. However, advancements in refractive surgery and contact lenses have provided additional options for addressing refractive errors. These methods, although effective, can be considered obtrusive since they require the individual to wear eyewear or undergo surgical procedures. This can lead to discomfort and, in rare cases, potential risks.

A recent addition to the options for correcting optical aberrations in the eye is the introduction of computational displays\cite{Authors6}. Unlike traditional methods that physically modify the eye’s refractive power through eyewear or surgery, computational displays\cite{Authors4, Authors5} present pre-distorted content based on the viewer’s prescription. This approach allows a sharp image to be formed on the retina without the need to alter the eye’s physical optics as the prescription changes. Computational-based vision correction displays \cite{Authors4, Authors5} do not require direct contact with the observer and can be built using readily available hardware components. Once the display device is constructed, digital correction of refractive errors is possible, eliminating the need for further adjustments to the optical hardware component.

Light field displays\cite{Authors3}, mainly used to display glasses-free 3D images, are now being explored for correcting visual aberrations in observers. These displays have the potential to revolutionise visual experiences by offering high-quality and personalised solutions in various modes, including 2D, glasses-free 3D, and vision correction. As technology continues to evolve, we can anticipate further improvements in resolution, contrast, and overall performance of light field displays.

In this paper, we report the simulations of a vision correction display solution that leverages the shearing and refraction of a light field as it propagates from the display to retina through the lens. Our approach involves utilising two methods for creating a light field display: modifying the LCD display with a lenslet array and with a pinhole array. With these techniques, we aim to improve the visual experience for individuals with various refractive errors. This innovative approach holds promise for advancing vision correction technology and enhancing accessibility for those with visual impairments. These ideas are present in \cite{Authors1}, and our contribution is in developing a simulation which gives the user the flexibility in designing such a system. In our understanding this is the first such simulator.

\subsection{Problem Definition}

This paper aims to create a simulation of a vision correction display (VCD) to enhance visual activities for visually impaired individuals. In this work, we only focus on the rectification of hyperopia and myopia.

\section{Display Light Field Formation}

In this work, we are generating display light field using the Inverse Light Field Projection which is introduced  in \cite{Authors1}. 
Light field\cite{Authors7,Authors8} is the most general representation of light  propagation. The plenoptic function\cite{Authors7}, regarded as the most general form of the light field, is a 7D function that describes all characteristics of light rays. However, this 7D plenoptic function can be reduced to a 4D representation using the two-plane model\cite{Authors2}. In this work, we focus on a four-dimensional light field with positions(2D) and directions(2D).


An image formed in the retina can be expressed as the integral of the light field inside the eye \cite{Authors1},
\begin{equation} \label{eq:1}
 I_r(x) = \int_{-\infty}^{\infty} L_r (x, u)A(u) \,du \ ,
\end{equation}
where $I_r$ is the image formed in the retina, $L_r$ is the retinal light ﬁeld, and $A$ is the aperture function, $x$ is a position on retina and $u$ is in aperture. \\

\begin{figure}
    \centering 
    \includegraphics[width=0.8\linewidth ]{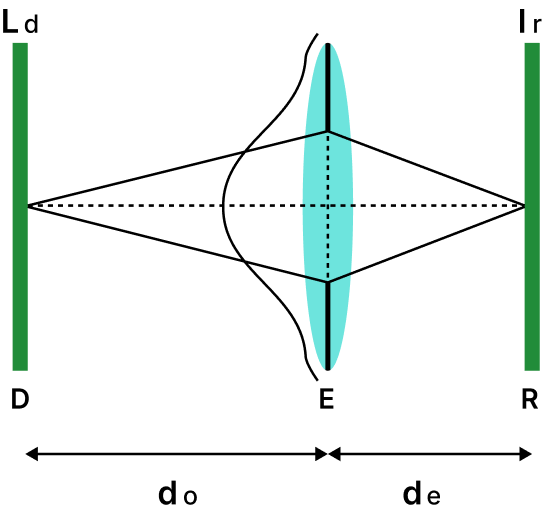}
    \caption{Image formation in retina from the display light field}
    \label{fig:1}
\end{figure}

The retinal light field depends on the light field from the display and can be calculated as explained below\cite{Authors1}. The display light field $L_d$, travels a distance $d_o$ to reach the pupil, resulting in a shearing of the light field in the x-direction, where the amount of shear depends on $d_o$. Subsequently, it undergoes refraction by the lens with a focal length $f$, causing a shearing in the y-direction with a shear dependence of $-1/f$. The light field then travels an additional distance $d_e$ within the eye until it reaches the retina. At the retina, the radiances are integrated to derive the retinal light field. Linear transforms on the light field can be used to depict propagation and refraction. A composite transform M can be employed to establish the relationship between the display light field $L_d$ and the retinal light field $L_r$ as follows.

\begin{align}\label{eq:2}
    \begin{pmatrix}
    x \\ u
    \end{pmatrix}  &= Q(d_e) T(d_e) R(f) T(d_o)\begin{pmatrix}
    x^d \\ u^d
    \end{pmatrix} \nonumber\\ 
   &=\begin{pmatrix}  1 &0 \\ 1& -d_e\end{pmatrix} 
    \begin{pmatrix}  1 &d_e \\ 0& 1\end{pmatrix} 
    \begin{pmatrix}  1 &0 \\ -1/f& 1\end{pmatrix} 
    \begin{pmatrix}  1 &d_o \\ 0& 1\end{pmatrix} 
    \begin{pmatrix}
    x^d \\ u^d
    \end{pmatrix} \nonumber\\
     &= M \begin{pmatrix}
    x^d \\ u^d 
    \end{pmatrix} ,
\end{align}   
where $Q(d_e)$ is the re-parameterization matrix to ensure that angular plane lies on the pupil, $T(d_e)$ is the shear matrix introduce by the propagation of light with in the eye, $R(f)$ is the refraction  matrix, $T(d_o)$ is the shear due to the propagation of light from display to the eyes and $\begin{pmatrix}
    x^d \\ u^d 
    \end{pmatrix}$ are the display light field parameters. $M$ is always invertible. Having established the relation between $L_r$ and $L_d$ (equation \ref{eq:2}), the equation \ref{eq:1} can be rewritten as,
\begin{align}\label{eg:3}
I_r(x) &= \int_{-\infty}^{\infty} L_r (x, u)A(u) \,du  \nonumber\\ 
 &=\int_{-\infty}^{\infty} L_d(M^{-1}(x, u))A(u) \,du ,
\end{align}
where the display light field $L_d$ is created by inversely transforming the retinal light field $L_r$.
This integral is discretized to produce a linear forward model.
\begin{equation}
I_r = PL_d ,
\end{equation}
where $P$ is the prefiltering matrix, $L_d$ is display light ﬁeld  and $I_r$ is the retinal image. The light field $L_d$ is essential for generating $I_r$, the retinal image. In section \ref{sec:DIF}, we elaborate on the process of generating the light field ($L_d$), which is created using the LCD display image $I_d$ with either a lenslet array or a pinhole array.

\begin{figure*}[h!t]
    \centering 
    \includegraphics[width=0.9\linewidth ]{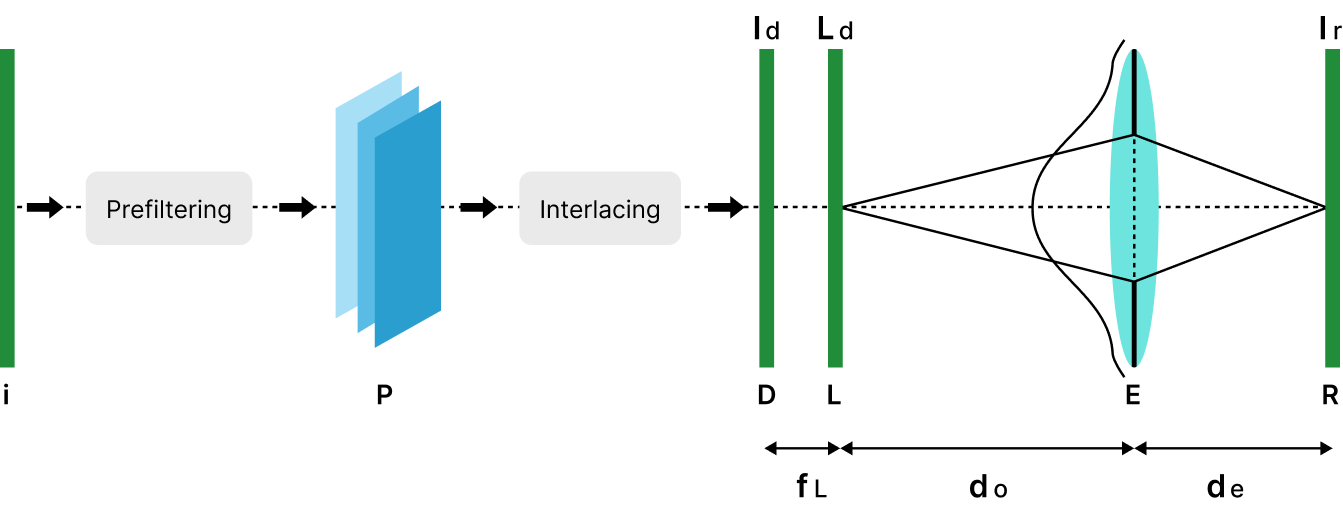}
    \caption{Formation of vision correction display and image formation in retina.}
    \label{fig:setup}
\end{figure*}

The goal of the vision correction display\cite{Authors1} is to provide the viewer with a 4D light field which produces the required 2D retinal projection. The following objective function can be optimised to determine the emitted light field if viewing distance, pupil size, and other factors are known. 
    \begin{align}
        & \argminB_{L_d}\quad || I_r- PL_d ||^2  \\
        & \text{subject to}\quad 0 \leq L_d \leq 1 \nonumber
    \end{align}
The optimized light field $L_d$ is refereed as the prefiltered light field which is capable of projecting correct image in the retina.\\

\subsection{Display Image formation}
\label{sec:DIF}
In this section the generation of $L_d$(the light field needed by the eye to perceive correct image) using a LCD with a pinhole array and LCD with a lenslet array are explained.

The pinhole array selectively permits the passage of pixel intensities from the pixels directly beneath each pinhole. Consequently, in order to construct a comprehensive light field using an LCD with a pinhole array, it is imperative to position the pixels of a single image at a distance such that each of them aligns directly under the consecutive pinholes. Failure to do so may result in certain pixel values being obstructed by the pinhole array, leading to the loss of critical information. Resolution of the vision correction display is very less compared to LCD display due to this behaviour of pinhole array.
In this case, the comprehensive light field will only be visible when then the optical axis of the eye and the axis of the pinhole array aligns. 

 The light rays originating from the pixels positioned under a single lens undergo refraction only by that specific lens for a given viewing angle. In the context of a lenslet array, the pixels situated under each lens for a given viewing angle are treated as a unified entity. Pixels under each such unit together forms a single light field segment. Thus, the combination of all such units collectively forms a comprehensive light field. For different viewing angles, the pixels positioned beneath each lens may differ, resulting in the creation of distinct light fields. As the pixels under each lens collectively constitute a single light field segment, this diminishes the resolution of the VCD significantly compared to the LCD used.

 For precise visualisation of the comprehensive light field of an image using an LCD with a lenslet array, it's crucial to position pixels from the same image consistently in same location beneath different lenses. Specifically, when the viewing angle aligns with the optical axis of the lenslet, the pixels belonging to the same image should be positioned at the centre of each respective lens in order to generate a comprehensive light field for that particular image.

Viewing angle significantly influences the accurate generation of the correct light field in both scenarios. Even slight variations in the viewing angle can result in incomplete visualisation of the light field. To address this challenge, shifted images of the same image are pre-filtered and subsequently interlaced.

Figure \ref{fig:setup} illustrates the entire system of the vision correction display. The workflow is as follows: Firstly, the image and its shifted images undergo pre-filtering, followed by interlacing to create the display image ($I_d$), which is then presented on the LCD (D). A lenslet array or pinhole array is positioned in front of the LCD at a distance equal to its focal length ($f_l$) to produce the light field ($L_d$). These light field rays propagate towards the eyes, where they are refracted by the eye's lens, forming an image on the retina. In cases of defocused eyes, a pre-filtered light field will be generated on the display, resulting in the production of a clear image for the eyes.


\section{Simulation of primitives}
In this study, we replicate the complete vision correction display \cite{Authors1} setup with Blender\cite{cite3B} software. The simulation involves representing the preprocessed image alongside either a pinhole array or a lenslet array, which serves as the vision correction display. Additionally, we employ a defocused camera within the Blender simulation to emulate the characteristics of a defocused human eye.
\subsection{Pinhole Array}
The pinhole array consists of evenly spaced pinholes arranged in both horizontal and vertical directions. During simulation, it is crucial to ensure that the pinholes are evenly positioned based on the spatio-angular trade-off ratio required.
The spatio-angular trade-off ratio is determined by the ratio between the distance separating the centres of the pinholes and the width of the pinhole. In our scenario, we consider a pinhole size of $100 \mu m$ and a distance of $500 \mu m$ between pinholes, resulting in a spatio-angular trade-off ratio of 5:1. Additionally, the focal length of the array is considered to be $3 mm$. Due to the limited amount of light rays passing through pinholes, it's essential to maintain high illumination of the display to achieve a bright light field in the VCD system with pinhole. In all simulations utilising a pinhole setup, the display emission strength is maintained at 50.

\begin{figure*}
  \centering
  \begin{subfigure}{0.22\linewidth}
    \includegraphics[width=\linewidth]{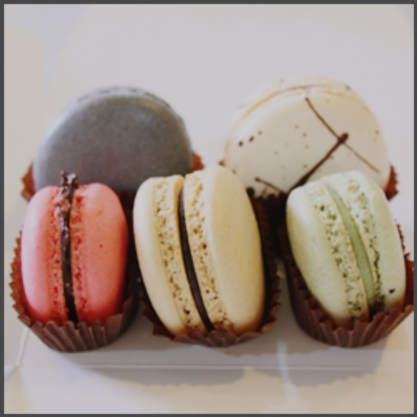}
    \caption{Original Image}
    \label{fig:3-a}
  \end{subfigure}
  \hfill
  \begin{subfigure}{0.22\linewidth}
    \includegraphics[width=\linewidth]{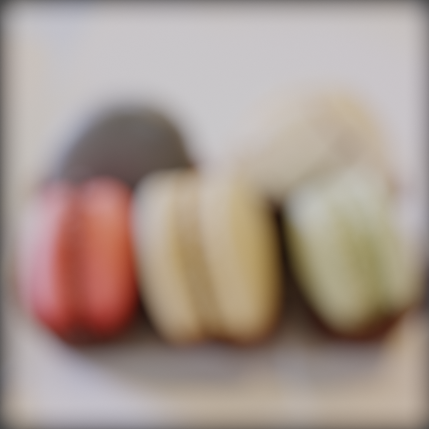}
    \caption{Defocused Image}
    \label{fig:3-b}
  \end{subfigure}
   \hfill
  \begin{subfigure}{0.22\linewidth}
    \includegraphics[width=\linewidth]{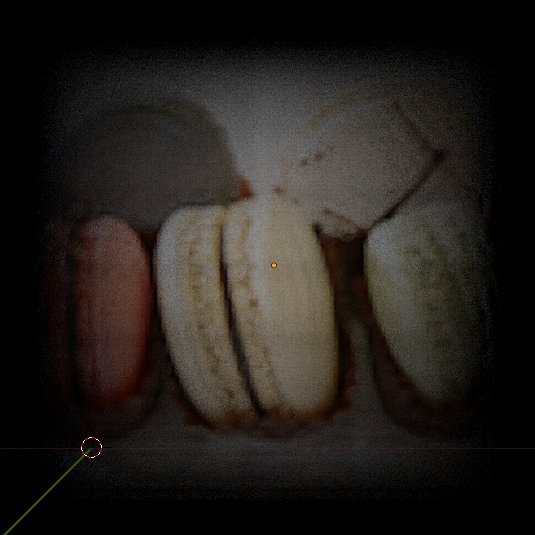}
    
    \caption{VCD image with pinhole array}
    \label{fig:3-c}
  \end{subfigure}
   \hfill
  \begin{subfigure}{0.22\linewidth}
    \includegraphics[width=\linewidth]{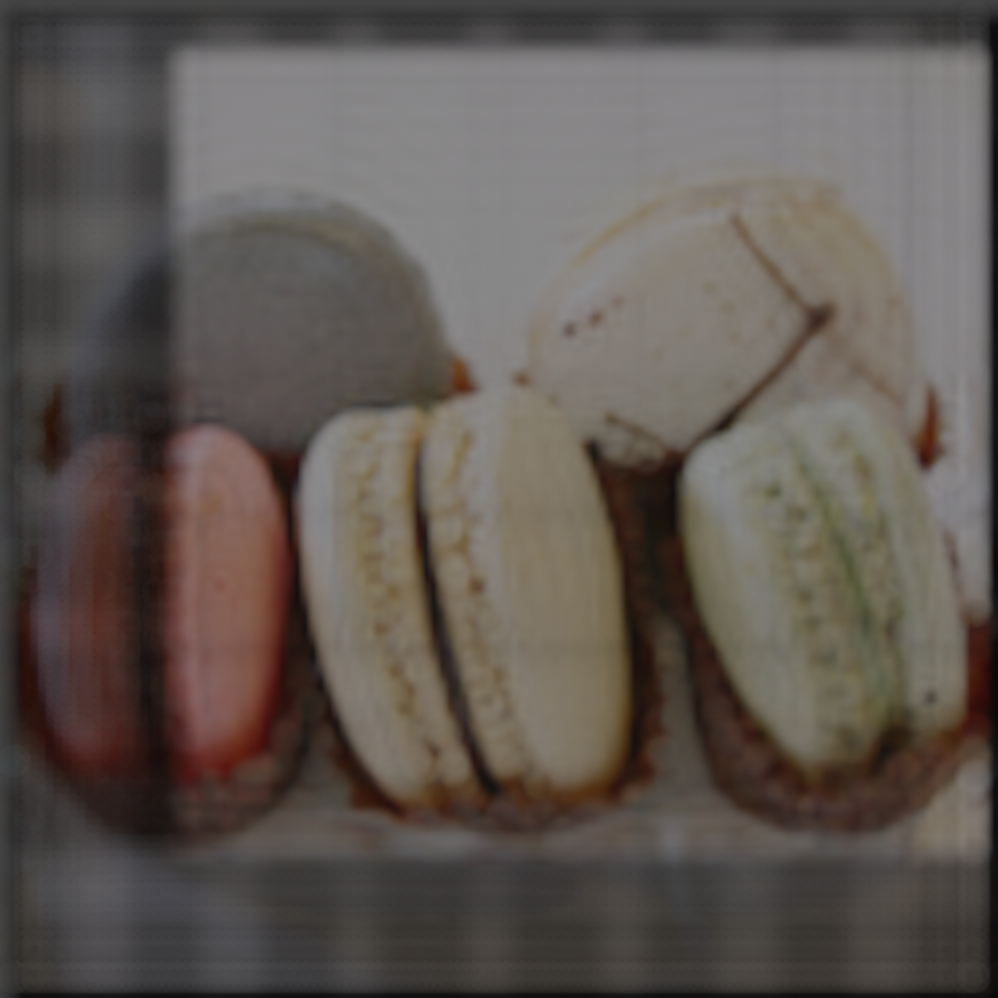}
    
    \caption{ VCD image with lenslet array}
    \label{fig:3-d}
  \end{subfigure}
  \caption{Output of the simulated hyperopic system.}
  \label{fig:3}
\end{figure*}

\begin{figure*}
  \centering
  \begin{subfigure}{0.22\linewidth}
    \includegraphics[width=\linewidth]{focused_h_25_38-crop.png}
    \caption{Original Image}
    \label{fig:4-a}
  \end{subfigure}
  \hfill
  \begin{subfigure}{0.22\linewidth}
    \includegraphics[width=\linewidth]{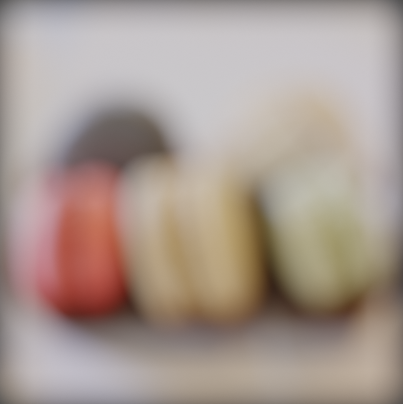}
    \caption{Defocused Image}
    \label{fig:4-b}
  \end{subfigure}
   \hfill
  \begin{subfigure}{0.22\linewidth}
    \includegraphics[width=\linewidth]{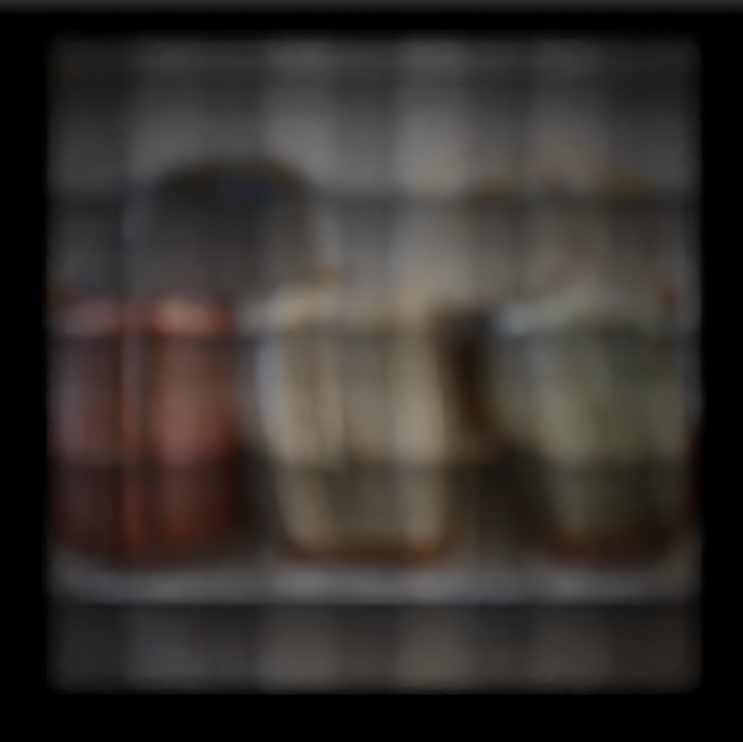}
    
    \caption{VCD image with pinhole array}
    \label{fig:4-c}
  \end{subfigure}
   \hfill
  \begin{subfigure}{0.22\linewidth}
    \includegraphics[width=\linewidth]{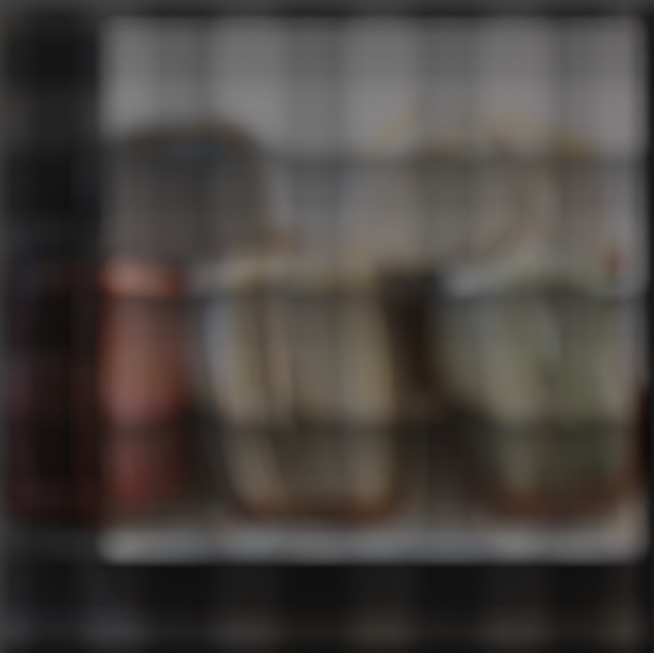}
    
    \caption{ VCD image with lenslet array}
    \label{fig:4-d}
  \end{subfigure}
  \caption{Output of the simulated myopic system.}
  \label{fig:4}
\end{figure*}

\subsection{Lenslet}
A lenslet array consists of a densely packed arrangement of micro lenses. Multiple types of lenslet arrays can be found in the market\cite{Cite1L, Cite2L}, each showcasing micro lenses with distinct shapes and configurations.
In this paper, we are using an array of square shaped micro lens with $3mm$ focal length $500\mu m$ lens diameter and $1 mm$ thickness as our lenslet array. We are opting for square lenslet array because of the easiness to interlacing the images and also due to the close packing nature of square lenslets. Standard illumination of the display screen is sufficient for the VCD system with lenslet array.

\subsection{Defocused Eye}

In our simulation, we model a fixed-focus camera to represent a defocused eye. For hyperopia, we define the fixed focal plane of the camera as the nearest focal plane of the eye. Additionally, we position the display closer to the eye than the fixed focal length to mimic a hyperopic eye.
For myopia, we consider the fixed focal plane of the camera as the farthest focal plane of the eye. Furthermore, we place the display image at a greater distance from the eye to simulate a myopic eye.

 \section{Simulations and Results}
We employ a camera equipped with a $50 mm$ lens set at $f/8$ for our simulations, representing a 6mm human pupil. And, the display used in all simulations features a pixel density of 254 pixels per inch (PPI). For both hyperopic and myopic systems, we conduct two simulations each, utilising both pinhole arrays and lenslets.

\subsection{Hyperopic system}

To simulate a hyperopic system, we focus the camera at a distance of $38 cm$ and position the display $25 cm$ away. The pinhole/lenslet array is placed $3 mm$ in front of the display. Detailed results are presented in Figure \ref{fig:3}. 
Figure \ref{fig:3-a} is the original image that is prefiltered for the hyperopic system. Figure \ref{fig:3-b} shows the image formed in retina by a simulated hyperopic system when the original image is displayed in the plain LCD setup, which is blurred as expected. The corrected image produced by the VCD system with pinhole array is shown in Figure \ref{fig:3-c}. This is the image formed in the retina when the prefiltered light field is given as input to the VCD system. The output image of the VCD system with pinhole array shows a  significant improvement in sharpness compared to the defocused blurred image. However, there is noticeable vignetting effect present at the edges of the output image. 
The vision corrected image produced by the VCD system with a lenslet array is shown in Figure \ref{fig:3-d}. This is the image formed in the retina of the hyperopic eye when the prefiltered light field is given as input to a VCD system with lenslet array. This image exhibits enhanced sharpness and clarity in comparison to the defocused image. Additionally, unlike the output image of the VCD system with pinhole array, the VCD system with lenslet array does not suffer from vignetting effect and appears brighter overall.

\subsection{Myopic System}
To simulate a myopic system, we focus the camera at a distance of $38 cm$ and position the display $25 cm$ away. The pinhole/lenslet array is placed $3 mm$ in front of the display. Detailed results are presented in Figure \ref{fig:4}.
Figure \ref{fig:4-a} is the original image used for prefiltering in the myopic system. Figure \ref{fig:4-b} shows the image formed in retina by a simulated myopic system when the original image is displayed in the plain LCD setup, which is blurred as expected. 
The corrected image produced by the VCD system with pinhole is shown in Figure \ref{fig:4-c}. This is the image formed in the retina of a myopic eye when prefiltered light field is given as input to a VCD system with pinhole array. The sharpness of this image is indeed improved in comparison to the defocused image. However, there are some vignetting effects and other artefacts are present in the output of the VCD system with pinhole array.
The vision corrected image produced in by VCD system with lenslet array is shown in Figure \ref{fig:4-d}. This is the image formed in the retina when prefiltered light field is given as input to a VCD system with lenslet array. The image formed in myopic eye by a  VCD with lenslet is sharper than image with plain LCD. But there are some artefacts are present the in the VCD output. 

\section{Discussion}

We have demonstrated a simulation for vision correction display with the purpose of providing a design tool for such displays. The initial results shown for hyperopia and myopia are promising. We are working on improving the myopia results and also on correcting the vignetting and removing other artefacts. We intend to extend these simulations to include other visual abberations too.

{\small
\bibliographystyle{ieee_fullname}
\bibliography{egbib}
}

\end{document}